\documentclass[
  preprint,onecolumn,a4paper,12pt,numerical,
  nofootinbib,longbibliography,floatfix
]{revtex4-2}

\newif\ifauthorversion
\authorversiontrue

\usepackage{multirow}
\usepackage{graphicx}
\usepackage{epstopdf}
\usepackage[caption=false,font=footnotesize]{subfig}
\usepackage{amsmath,amsfonts,amssymb,bm,upgreek,mathrsfs,mathcomp}
\usepackage{booktabs}
\usepackage{threeparttable}
\usepackage{siunitx}
\usepackage[dvipsnames]{xcolor}
\usepackage[
  colorlinks=true,
  citecolor=blue,
  urlcolor=blue,
  breaklinks=true,
  pdfauthor={},
  pdftitle={Prospects for Precision YN and anti-n N Measurements with an LH2/LD2 Target in the BESIII Detector},
  pdfsubject={},
  pdfkeywords={hyperon-nucleon scattering, antineutron-nucleon scattering, cryogenic internal target, effective luminosity, tracking performance, tau-charm factory}
]{hyperref}

\graphicspath{{figures/}}
\renewcommand{\arraystretch}{1.1}

\setcitestyle{numbers,square,sort&compress}

\begin{document}
\makeatletter\renewcommand{\p@subsection}{}\renewcommand{\p@subsubsection}{}\makeatother

\title{\texorpdfstring{Prospects for Precision $YN$ and $\bar{n}N$ Measurements with an LH$_2$/LD$_2$ Target in the BESIII Detector}{Prospects for Precision YN and anti-n N Measurements with an LH2/LD2 Target in the BESIII Detector}}

\ifauthorversion
  \author{Zhao-Ling Zhang$^{a}$, Xu Gao$^{a}$, Wei-Min Song$^{a}$, and Chang-Zheng Yuan$^{b,c}$}
\affiliation{$^{a}$College of Physics, Jilin University, Changchun 130012, China\\
$^{b}$Institute of High Energy Physics, Chinese Academy of Sciences, Beijing 100049, China\\
$^{c}$University of Chinese Academy of Sciences, Beijing 100049, China}

\fi

\begin{abstract}
Tagged hyperon and antineutron beams from $J/\psi$ and $\psi(3686)$ decays allow BESIII to study (anti)hyperon--nucleon ($YN$) and antineutron--nucleon ($\bar{n}N$) scattering, but the present measurements use the beam pipe as the target, whose only free protons are in a thin cooling-oil layer. We present a conceptual feasibility study of a liquid hydrogen (LH$_2$) or liquid deuterium (LD$_2$) target between the beam pipe and the Cylindrical Gas Electron Multiplier Inner Tracker. Relative to the beam pipe, a 30~mm LH$_2$ layer increases the effective free-proton areal density by factors of 7--22 for $\Lambda$, $\Sigma^{+}$, $\Xi$, and $\bar n$ beams (3.4 for $\Omega^-$). A 30~mm LD$_2$ layer gives 0.9--2.9 times the effective neutron areal density for the same beams (0.4 for $\Omega^-$) and replaces the composite Be, C, and Au nuclei with the deuteron. These ratios characterize the target itself and do not include channel-dependent efficiencies, backgrounds, or deuteron corrections. In detector simulations without localized support material, the target affects tracking mainly below 0.5~GeV/$c$; above this momentum, proton, kaon, and pion efficiencies stay within 1.6 percentage points of the baseline. In a signal-only full-event simulation of $\bar np\to2\pi^+\pi^-$, the LH$_2$-to-beam-pipe selection-efficiency ratio is 0.86, and the vertex and energy-closure resolutions are comparable to those for beam-pipe interactions. An internal LH$_2$/LD$_2$ target is thus a promising concept for higher-statistics and less model-dependent $YN$ and $\bar nN$ measurements at BESIII, requiring further engineering and channel-level validation.
\end{abstract}

\keywords{hyperon-nucleon scattering, antineutron-nucleon scattering, cryogenic internal target, effective luminosity, tracking performance, tau-charm factory}

\maketitle

\noindent\textbf{Keywords:} hyperon-nucleon scattering; antineutron-nucleon scattering; cryogenic internal target; effective luminosity; tracking performance; tau-charm factory.
\par\medskip
\section{Introduction}

Baryon--baryon and baryon--antibaryon interactions offer fundamental tests of the strong interaction in the nonperturbative regime of quantum chromodynamics (QCD) and provide critical insights into the internal structure of hadrons~\cite{Epelbaum:2008ga,Machleidt:2011zz,Bogner:2009bt,Klempt:2002ap,Bressani:2003pv,Gal:2016boi}. While scattering experiments provide the most direct probe of these forces, experimental progress has long been hindered by severe beam limitations: hyperons decay within a few centimeters, and antineutrons, being neutral, cannot be accelerated or cooled, so antineutron beams are available only as low-intensity secondary beams.

Hyperon--nucleon ($YN$) interactions generalize nuclear forces to the strangeness sector. Beyond probing SU(3) flavor-symmetry breaking and dictating hypernuclear binding, they govern the properties of hyperonic matter inside neutron stars. Hyperons are predicted to appear in neutron-star cores at two to three times the nuclear saturation density, where they soften the equation of state (EOS)---a softening that generally conflicts with the astronomical observations of two-solar-mass pulsars~\cite{Tolos:2020aln,Demorest:2010bx,Antoniadis:2013pzd}. Resolving this ``hyperon puzzle'', for instance via repulsive three-body $YNN$ forces~\cite{Lonardoni:2014bwa,Gerstung:2020ktv}, fundamentally requires high-precision constraints on the underlying two-body $YN$ potentials. Historically, however, $YN$ scattering data have been scarce, relying primarily on bubble-chamber measurements from the 1960s and 1970s~\cite{Alexander:1968xn,Sechi-Zorn:1968zln,Kadyk:1971tc,Eisele:1971mk,Hauptman:1977hr}. Although recent experiments at J-PARC~\cite{J-PARCE40:2021qtk,J-PARCE40:2021qxa,J-PARCE40:2022nvq} and CLAS~\cite{CLAS:2021kpz} have provided improved $\Sigma^\pm p$ and $\Lambda p$ data, the available kinematic coverage and statistical precision remain severely limited.

Consequently, modern $YN$ potentials are calibrated primarily in the low-momentum regime. Chiral effective field theory ($\chi$EFT) potentials describe the coupled $\Lambda N$--$\Sigma N$ system up to next-to-next-to-leading order~\cite{Haidenbauer:2023nnlo}, with covariant chiral formulations predicting cross sections up to $\sim 900$~MeV/$c$~\cite{Song:2022covariant}. At higher momenta, existing CLAS $\Lambda p$ data between 0.9 and 2.0~GeV/$c$~\cite{CLAS:2021kpz} already reveal substantial discrepancies, demonstrating the need for higher-order chiral potentials~\cite{Song:2022covariant}. In this transition region, precise differential cross sections are essential, as they can break theoretical degeneracies among models that agree on total cross sections~\cite{Song:2022covariant}. In the strangeness $S=-2$ sector, the $\Xi N$ interaction is even less constrained: lattice QCD near the physical point provides only low-energy $S$-wave predictions~\cite{Sasaki:2020lattice}, and experimental data on channels such as $\Xi^0 n \to \Xi^- p$ have only emerged very recently.

In parallel, the nucleon--antinucleon ($\bar{N}N$) sector provides an indispensable laboratory for understanding annihilation dynamics and meson spectroscopy~\cite{OBELIX:1998nvy,OBELIX:1994hbn}. In particular, the antineutron--proton ($\bar{n}p$) system represents a pure isospin $I=1$ state completely free of the Coulomb interactions present in $\bar{p}p$ scattering. Past measurements, such as those at BNL E-767~\cite{BNL} and OBELIX at CERN~\cite{Agnello:1997wq,OBELIX:2000kga,Astrua:2002zg}, used such beams produced via the charge-exchange reaction $\bar{p}p \to \bar{n}n$, and antineutron scattering data above 500~MeV/$c$ remain scarce.

A breakthrough approach to overcome these beam limitations was proposed in Ref.~\cite{Yuan:2021yks}: utilizing charmonium ($J/\psi$ and $\psi(3686)$) decays at high-luminosity $e^+e^-$ colliders as a prolific source of tagged hyperons and antineutrons~\cite{Dai:2024myk}. Due to large resonance production cross sections and sizable baryonic decay branching fractions, this method produces monoenergetic hyperon beams (such as $\Lambda$ at 1074~MeV/$c$ and $\Sigma^+$ at 992~MeV/$c$) as well as broad-spectrum antineutrons up to 1.17~GeV/$c$ (Table~\ref{tab:hyperon_decay_modes}). Fully reconstructing the recoiling partner baryon uniquely tags the incident projectile and determines its four-momentum from the two-body or three-body decay kinematics. BESIII has successfully demonstrated this concept by utilizing its existing beam pipe as an interaction target, producing pioneering measurements of $\bar{n}p$, $\Xi^0 n$, $\Lambda N \to \Sigma^+ X$, $\Lambda p$, $\bar{\Lambda} p$, and $\Sigma^+ n$ processes~\cite{BESIII:2025yup,BESIII:2023clq,BESIII:2023trh,BESIII:2024geh,BESIII:2025bft}.

Despite this success, beam-pipe scattering faces two intrinsic bottlenecks. First, free target protons exist only in the hydrogen atoms of a thin (0.8~mm) cooling-oil layer, which severely restricts the statistical precision and angular binning of differential cross-section measurements. Second, while the gold (Au) and beryllium (Be) layers and the carbon (C) of the cooling oil provide abundant nucleons, they are bound within nuclei; extracting fundamental, elementary baryon--nucleon cross sections from these targets requires substantial nuclear-model corrections, introducing irreducible systematic uncertainties.

Installing a dedicated, high-density cryogenic target can simultaneously resolve both limitations. While placing custom targets just outside the beam pipe was suggested in Ref.~\cite{Yuan:2021yks} and effective luminosities for existing detector materials were formulated in Ref.~\cite{Dai:2024myk}, this paper presents the first conceptual feasibility study of an internal liquid hydrogen (LH$_2$) or liquid deuterium (LD$_2$) target integrated directly into the BESIII detector. Located in the radial clearance between the beam pipe and the Cylindrical Gas Electron Multiplier Inner Tracker (CGEM-IT), an LH$_2$ target dramatically amplifies the free-proton areal density, while an LD$_2$ target replaces composite nuclear targets with the deuteron---providing a simple, weakly bound system for quasi-free neutron measurements.

The remainder of this paper is structured as follows. Section~\ref{sec:target_design} details the target cell configuration, the passive material budget, and the kinematics of the incident hyperon and antineutron beams. Section~\ref{sec:detector_performance} evaluates the impact of the added target material in the full BESIII Geant4 detector simulation, first on single charged tracks and then on the complete reconstruction of the exclusive channel $J/\psi \to p \pi^- \bar{n}$ with $\bar{n}p \to 2\pi^+ \pi^-$. Section~\ref{sec:signal_yields} quantifies the effective areal densities and presents projected annual signal yields and systematic uncertainty comparisons against existing beam-pipe measurements. Finally, Sec.~\ref{sec:summary} summarizes the findings and discusses the outlook for future facilities.

\section{Target configuration and incident beams}
\label{sec:target_design}

\subsection{Detector overview and target configuration}
\label{subsec:target_config}

The BESIII detector~\cite{BESIII:2009fln} is a magnetic spectrometer at the BEPCII $e^{+}e^{-}$ collider~\cite{Yu:2016cof}. A multilayer drift chamber (MDC), a time-of-flight system (TOF), and a CsI(Tl) electromagnetic calorimeter (EMC) are enclosed in a superconducting solenoid that provides a 1.0~T magnetic field, and a muon chamber system (MUC) is embedded in the return yoke. The CGEM-IT replaces the inner part of the MDC~\cite{Balossino:2022ywn}.

The proposed target is an annular cryogenic cell in the radial gap between the outer surface of the beam pipe ($r = 33.7$~mm) and the inner radius of the CGEM-IT ($r = 76.9$~mm)~\cite{Balossino:2022ywn}. In the nominal configuration, the liquid layer is 30~mm thick ($40.4 < r < 70.4$~mm) and extends over $|z| < 200$~mm, symmetric about the interaction point, so that it covers the full $|\cos\theta| < 0.8$ acceptance used in Sec.~\ref{sec:signal_yields}; together with its walls and insulation, it lies within $35 < r < 76$~mm.

\begin{figure*}[t]
  \centering
  \includegraphics[width=0.95\textwidth]{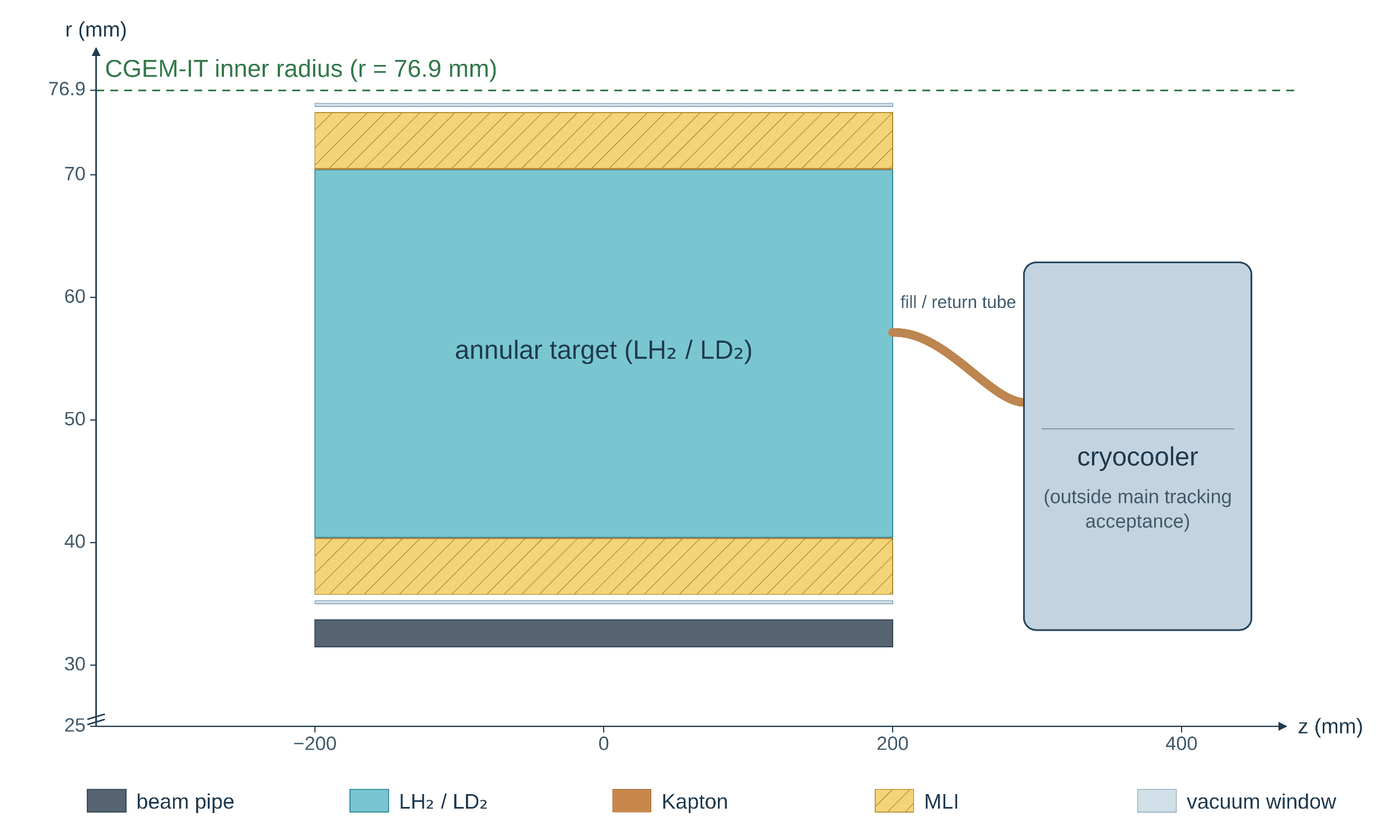}
  \caption{Axial ($r$--$z$) schematic of the proposed BESIII cryogenic target.
    The annular LH$_2$/LD$_2$ volume between the beam pipe ($r = 33.7$~mm) and the CGEM-IT inner radius ($r = 76.9$~mm) is bounded by inner and outer Kapton walls and surrounded by multi-layer insulation and Mylar--aramid vacuum windows.
    A representative fill/return line and the external cryocooler are drawn schematically outside the main tracking acceptance.}\label{fig:cryotarget_schematic}
\end{figure*}

The liquid is contained between inner and outer Kapton walls, each 0.1~mm thick. Kapton is a low-$Z$ material ($Z_{\text{eff}} \sim 5$) that tolerates radiation and remains mechanically robust at cryogenic temperatures. To limit radiative heat transfer, the cell is wrapped in a 10-layer aluminized-Mylar multi-layer insulation (MLI) blanket inside an evacuated insulation space closed by vacuum windows of Mylar laminated on aramid fabric. These materials are taken from the MUSE liquid-hydrogen target~\cite{Roy:2019add}, which uses a 120~$\mu$m Kapton cell wall, 10 layers of aluminized-Mylar superinsulation, and 368~g/m$^2$ Mylar--aramid vacuum windows. Table~\ref{tab:passive_material} lists the layers in radial order, and Fig.~\ref{fig:cryotarget_schematic} shows an $r$--$z$ section of the target. As a candidate cooling scheme, an external Gifford--McMahon (GM) cryocooler would keep the liquid near its normal boiling point; its cold head and condenser would sit outside the main tracking acceptance, connected to the cell by a low-mass fill/return line.

At their normal boiling points of 20.3 and 23.7~K, liquid hydrogen and deuterium have densities of 0.0708 and 0.164~g\,cm$^{-3}$. LH$_2$ provides free protons for $Yp$ and $\bar{n}p$ scattering, so the signal interaction needs no bound-nucleon correction. LD$_2$ provides neutrons for $Yn$ scattering with the crucial advantage of a much simpler and cleaner nuclear environment: each neutron is accompanied only by a single proton bound by 2.22~MeV, so that deuteron corrections, which can be treated as a few-body problem, replace the model-dependent treatment of the composite Be, C, and Au nuclei.

\begin{table}[t]
\caption{
Continuous passive-material layers of the nominal 30~mm target, in radial order, as included in the detector simulation of Sec.~\ref{sec:detector_performance}.
}
\label{tab:passive_material}
\begin{ruledtabular}
\begin{tabular}{ll}
Component & Material and thickness \\ \hline
Inner vacuum window & Mylar--aramid laminate, $368~\mathrm{g/m^2}$ (simulated as $0.26~\mathrm{mm}$ Mylar) \\
Inner MLI blanket & Al-Mylar, $10\times0.006~\mathrm{mm}$ \\
Inner target wall & Kapton, $0.1~\mathrm{mm}$ \\
Liquid target & LH$_2$ or LD$_2$, $30~\mathrm{mm}$ \\
Outer target wall & Kapton, $0.1~\mathrm{mm}$ \\
Outer MLI blanket & Al-Mylar, $10\times0.006~\mathrm{mm}$ \\
Outer vacuum window & Mylar--aramid laminate, $368~\mathrm{g/m^2}$ (simulated as $0.26~\mathrm{mm}$ Mylar) \\
\end{tabular}
\end{ruledtabular}
\end{table}

\subsection{Incident beams and survival probabilities}
\label{subsec:incident_beams}

The laboratory decay lengths $\lambda_Y=(p_Y/m_Y)\,c\tau_Y$ of the hyperons considered here, where $p_Y$, $m_Y$, and $\tau_Y$ are the momentum, mass, and lifetime, range from 1.1~cm for the $\Omega^-$ to 7.6~cm for the $\Lambda$, comparable to the radii of the target, so a sizable fraction decays before reaching the liquid or while crossing it. The flux reaching the target therefore depends on the momentum, lifetime, and angular distribution of each species, which we specify first.

The hyperons are produced in the two-body decays $J/\psi \to Y\bar{Y}$ for the $\Lambda$, $\Sigma^+$, and $\Xi$, and $\psi(3686) \to \Omega^-\bar{\Omega}^+$ for the $\Omega^-$, whose pair production is kinematically forbidden in $J/\psi$ decays. In the parent rest frame, each hyperon has a fixed momentum $p_Y=\sqrt{m_{\mathrm{parent}}^2/4-m_Y^2}$ and the angular distribution $f_Y(\cos\theta)\propto1+\alpha\cos^2\theta$~\cite{Faldt:2017kgy,Perotti:2018wxm}, normalized to unity over $-1\le\cos\theta\le1$, where $\alpha$ is the angular-distribution parameter of the production process measured by BESIII~\cite{BESIII:2022qax,BESIII:2021ypr,BESIII:2020fqg,BESIII:2023drj,BESIII:2020lkm}. Table~\ref{tab:hyperon_decay_modes} lists these parameters together with the momenta, branching fractions, masses, and lifetimes~\cite{ParticleDataGroup:2024cfk}; the same inputs are used for the areal densities of Sec.~\ref{subsec:target_densities}. The small boost from the BEPCII beam-crossing angle is neglected for the hyperons. The angular acceptance $A_Y=\int_{-0.8}^{0.8}f_Y(\cos\theta)\,d\cos\theta$, the fraction of produced particles with $|\cos\theta|<0.8$, is 0.75--0.78 for the $\Lambda$, $\Xi$, and $\Omega^-$, whose positive $\alpha$ favors production at small polar angles, and 0.859 for the $\Sigma^+$, whose negative $\alpha = -0.508$ favors production near $90^\circ$.

Antineutrons from the three-body decay $J/\psi \to p\pi^{-}\bar{n}$ have a continuous momentum spectrum up to 1.17~GeV/$c$, and their angular distribution cannot be described by a single $1 + \alpha\cos^2\theta$ form. We therefore use $5\times10^5$ events generated with the partial-wave-analysis (PWA) model of the BESIII measurement~\cite{BESIII:2025yup}. Of these events, 415,250 satisfy $|\cos\theta_{\bar n}| < 0.8$, giving $A_{\bar n} = 0.8305 \pm 0.0005$.

\begin{figure}[t]
  \centering
  \includegraphics[width=0.55\textwidth]{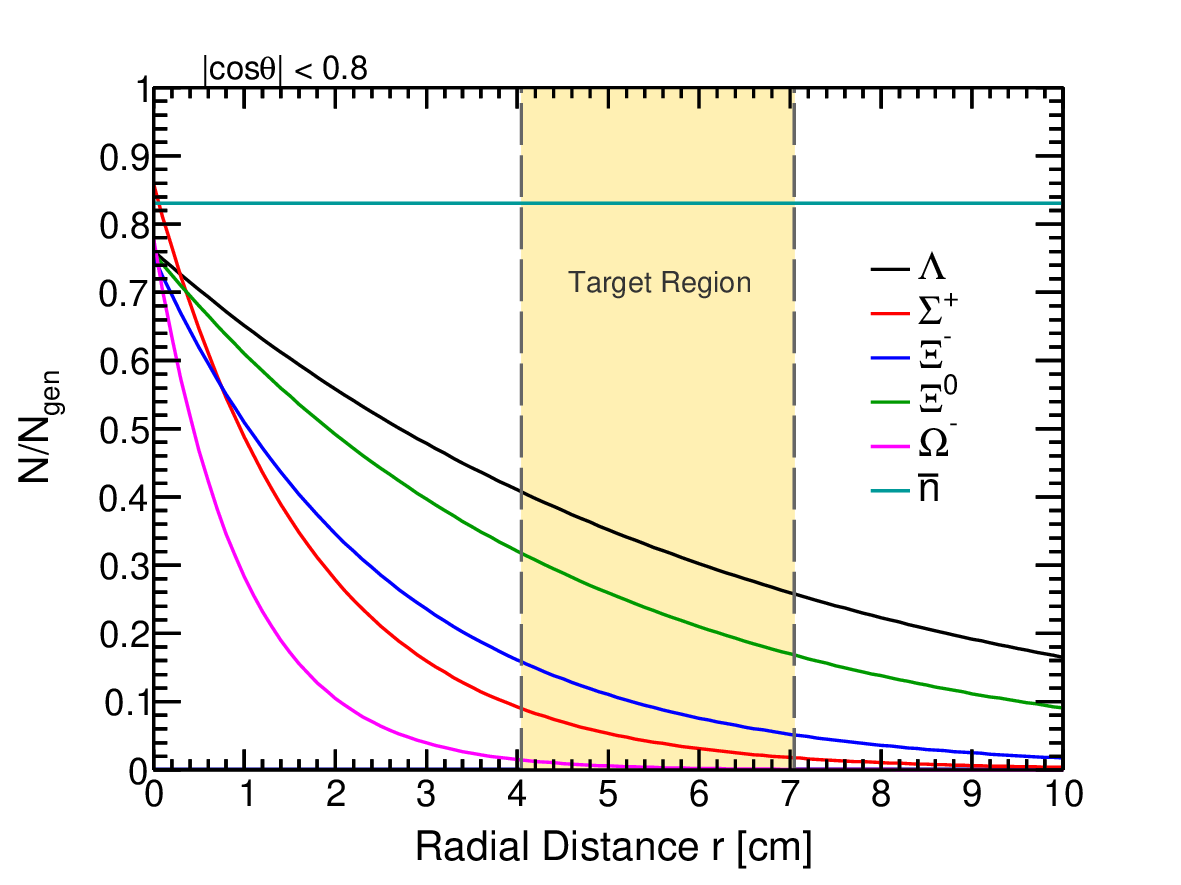}
  \caption{Accepted surviving fraction $N/N_{\mathrm{gen}}$ of $\Lambda$, $\Sigma^{+}$, $\Xi^{-}$, $\Xi^{0}$, $\Omega^{-}$, and $\bar{n}$ as a function of the radial distance $r$ from the interaction point, computed from Eq.~\eqref{eq:decay_integration} averaged over $|\cos\theta|<0.8$; the value at $r=0$ equals the angular acceptance $A_Y$. The yellow band marks the target region. The antineutron line is the PWA acceptance $A_{\bar n}=0.83050$.}
  \label{fig:survival_factor}
\end{figure}

Following Ref.~\cite{Dai:2024myk}, a particle emitted at polar angle $\theta$ travels a path $r/\sin\theta$ to reach radius $r$ and survives with probability
\begin{equation}
S_Y(r,\theta) = \exp\!\left(-\frac{r}{\lambda_Y\sin\theta}\right).
\label{eq:decay_integration}
\end{equation}
The accepted surviving fraction is its average over the accepted polar angles,
\begin{equation}
\frac{N(r)}{N_{\mathrm{gen}}}=\int_{-0.8}^{0.8} f_Y(\cos\theta)\,S_Y(r,\theta)\,d\cos\theta,
\label{eq:survival_fraction}
\end{equation}
where $N_{\mathrm{gen}}$ is the number of particles produced at the interaction point and $N(r)$ is the number that reach $r$ within $|\cos\theta|<0.8$; at $r=0$ it equals $A_Y$. For the antineutron, $S_{\bar n}=1$ and the fraction is the constant $A_{\bar n}$. Figure~\ref{fig:survival_factor} shows the resulting curves.

\begin{table*}[t]
  \centering
  \caption{Production channels and inputs for the incident species. Branching fractions (BF), masses, and lifetimes are from PDG 2024~\cite{ParticleDataGroup:2024cfk}; $p$ is the momentum in the parent rest frame. The two errors on $\alpha$ are statistical and systematic; the $\Omega^-$ value has a combined error. The antineutron kinematics are taken from the PWA sample described in the text.}
  \label{tab:hyperon_decay_modes}
  \scriptsize
  \setlength{\tabcolsep}{3pt}
  \begin{tabular}{lllllll}
  \toprule
  Species & Production & BF ($10^{-3}$) & $p$ (MeV/$c$) & $m_Y$ (MeV/$c^2$) & $\tau_Y$ ($10^{-10}$~s) & $\alpha$ \\
  \midrule
  $\Lambda$ & $J/\psi\to\Lambda\bar{\Lambda}$ & $1.89\pm0.09$ & 1074 & $1115.683\pm0.006$ & $2.617\pm0.010$ & $0.4748\pm0.0022\pm0.0031$~\cite{BESIII:2022qax} \\
  $\Sigma^{+}$ & $J/\psi\to\Sigma^{+}\bar{\Sigma}^{-}$ & $1.07\pm0.04$ & 992 & $1189.37\pm0.07$ & $0.8018\pm0.0026$ & $-0.508\pm0.006\pm0.004$~\cite{BESIII:2020fqg} \\
  $\Xi^{-}$ & $J/\psi\to\Xi^{-}\bar{\Xi}^{+}$ & $0.97\pm0.08$ & 807 & $1321.71\pm0.07$ & $1.639\pm0.015$ & $0.586\pm0.012\pm0.010$~\cite{BESIII:2021ypr} \\
  $\Xi^{0}$ & $J/\psi\to\Xi^{0}\bar{\Xi}^{0}$ & $1.17\pm0.04$ & 818 & $1314.86\pm0.20$ & $2.90\pm0.09$ & $0.514\pm0.006\pm0.015$~\cite{BESIII:2023drj} \\
  $\Omega^{-}$ & $\psi(3686)\to\Omega^{-}\bar{\Omega}^{+}$ & $0.056\pm0.003$ & 774 & $1672.45\pm0.29$ & $0.821\pm0.011$ & $0.24\pm0.10$~\cite{BESIII:2020lkm} \\
  $\bar{n}$ & $J/\psi\to p\pi^{-}\bar{n}$ & $2.12\pm0.09$ & 0--1174 & -- & stable & PWA sample \\
  \bottomrule
  \end{tabular}
\end{table*}

\section{Detector simulation and reconstruction performance}
\label{sec:detector_performance}

\subsection{Simulation setup}
\label{subsec:sim_setup}

The introduction of target material between the beam pipe and the CGEM-IT inevitably impacts charged-particle tracking via energy loss, multiple Coulomb scattering, and secondary hadronic interactions. Following the methodology in Ref.~\cite{Yuan:2024gyl}, we investigate these effects using the full BESIII detector simulation, systematically comparing the baseline detector (without target) against the nominal 30~mm LH$_2$ and LD$_2$ configurations, each incorporating the full passive material budget listed in Table~\ref{tab:passive_material}. Localized mechanical supports, alignment spacers, and cryogenic service feedthroughs are omitted from the present model.

Both event simulation and track reconstruction are performed within the BESIII Offline Software System (BOSS), version 7.2.0, utilizing the official geometry description with the CGEM-IT included~\cite{BESIII:2009fln,Balossino:2022ywn}. Particle transport is modeled with \textsc{Geant4} (version 10.7.p02) using the \texttt{FTFP\_BERT} physics list. BOSS 7.2.0 is a development release with preliminary CGEM-IT simulation and reconstruction, so absolute efficiencies and resolutions may differ from those of future production releases. We therefore compare each target configuration with the baseline using the same software version and simulation settings, and quote the changes relative to the baseline.

The performance evaluation relies on two complementary Monte Carlo samples:
first, single-particle samples ($p$, $\bar{p}$, $K^\pm$, $\pi^\pm$) generated at the interaction point are used to assess basic tracking efficiencies and track parameter resolutions across all configurations (Sec.~\ref{subsec:tracking}); second, a full-event exclusive sample of the reaction chain $J/\psi \to p\pi^-\bar{n}$ with $\bar{n}p \to 2\pi^+\pi^-$ is employed to benchmark complete multi-track vertexing and kinematic reconstruction for annihilations occurring in the beam pipe and in LH$_2$ (Sec.~\ref{subsec:nbarp_simulation}).

\subsection{Single-particle tracking performance}
\label{subsec:tracking}

The single-particle benchmark samples consist of protons, antiprotons, charged kaons, and charged pions generated at the nominal interaction point, with approximately $10^6$ events simulated per species and target configuration, divided equally between the two charge states. Momentum and $\cos\theta$ are generated uniformly. The performance is compared in truth-momentum bins of 0.1~GeV/$c$ over $0.3$--$1.5$~GeV/$c$ for protons and antiprotons, $0.2$--$1.5$~GeV/$c$ for kaons, and $0.1$--$1.5$~GeV/$c$ for pions, and in $|\cos\theta|$ bins of width $0.1$. Each momentum projection integrates over the full angular acceptance, while each angular projection integrates over the accessible momentum range for that species.

A reconstructed track candidate is accepted if the Kalman-filter fit successfully converges under the mass hypothesis of the generated particle. A candidate is matched to the primary truth track if more than $50\%$ of its associated MDC wire hits originate directly from that particle (excluding hits from secondary $\delta$-electrons); hits associated with other particles or carrying ambiguous truth status count against the match. The reconstruction efficiency is defined as
\begin{equation}
\epsilon_{\mathrm{rec}}=\frac{N_{\mathrm{match}}}{N_{\mathrm{gen}}},
\label{eq:eff_rec}
\end{equation}
where $N_{\mathrm{match}}$ denotes the number of generated particles with at least one matched candidate. If multiple candidates satisfy the matching criteria, the one possessing the greatest number of primary hits is chosen for evaluating track resolution.

Track parameter resolutions are evaluated via the relative momentum residual, $\delta_p=(p_{\mathrm{rec}}-p_{\mathrm{truth}})/p_{\mathrm{truth}}$, and the polar-angle residual, $\delta_\theta=\theta_{\mathrm{rec}}-\theta_{\mathrm{truth}}$. Their widths in each bin are determined from the unbinned distributions of matched tracks using the non-parametric estimator
\begin{equation}
\sigma_{68}(x)=\frac{q_{84.135}(x)-q_{15.865}(x)}{2},
\label{eq:sigma68}
\end{equation}
where $q_a$ is the $a$th percentile of the residual distribution, so that $\sigma_{68}$ is the half-width of its central 68.27\% interval. The efficiency uncertainties are 68.27\% Wilson score intervals for a binomial proportion~\cite{Wilson:1927zz}, which remain within $[0,1]$ and are reliable for efficiencies close to unity. The uncertainty of $\sigma_{68}$ is estimated with the nonparametric bootstrap~\cite{Efron:1979bxm}: the matched tracks in each bin are resampled with replacement 2000 times, $\sigma_{68}$ is recomputed for each resample, and the standard deviation of the 2000 values is quoted.

\begin{figure*}[t]
  \centering
  \includegraphics[width=\textwidth]{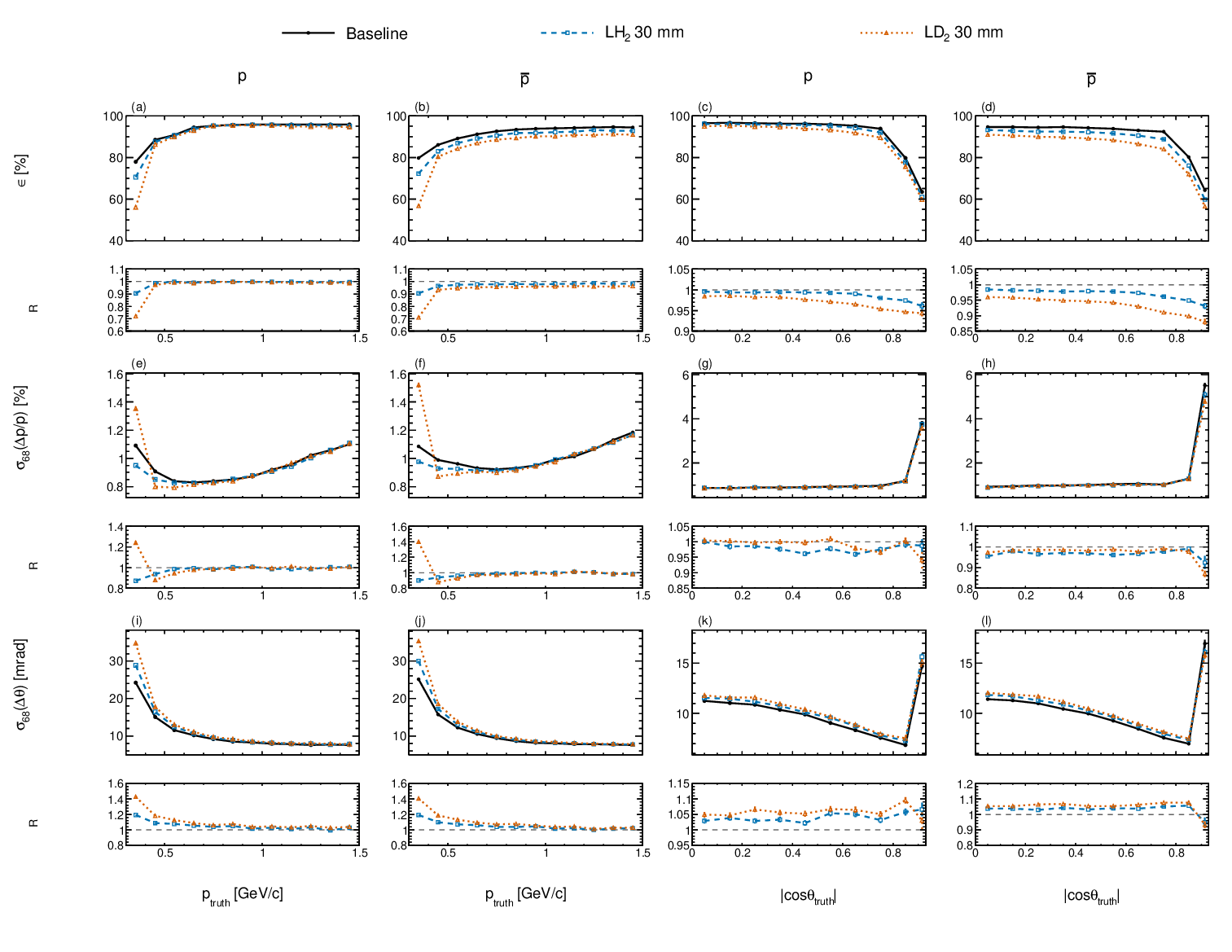}
  \caption{Single-particle tracking performance for protons and antiprotons. Rows show the reconstruction efficiency, relative momentum-residual width $\sigma_{68}(\delta_p)$, and polar-angle width $\sigma_{68}(\delta_\theta)$. Columns show distributions versus truth momentum for protons and antiprotons, followed by distributions versus $|\cos\theta_{\mathrm{truth}}|$. Black circles, blue squares, and orange triangles represent the baseline, 30~mm LH$_2$, and 30~mm LD$_2$ setups, respectively. The sub-panel below each plot displays the ratio $R$ of each target configuration to the baseline.}
  \label{fig:single_particle_proton}
\end{figure*}

\begin{figure*}[t]
  \centering
  \includegraphics[width=\textwidth]{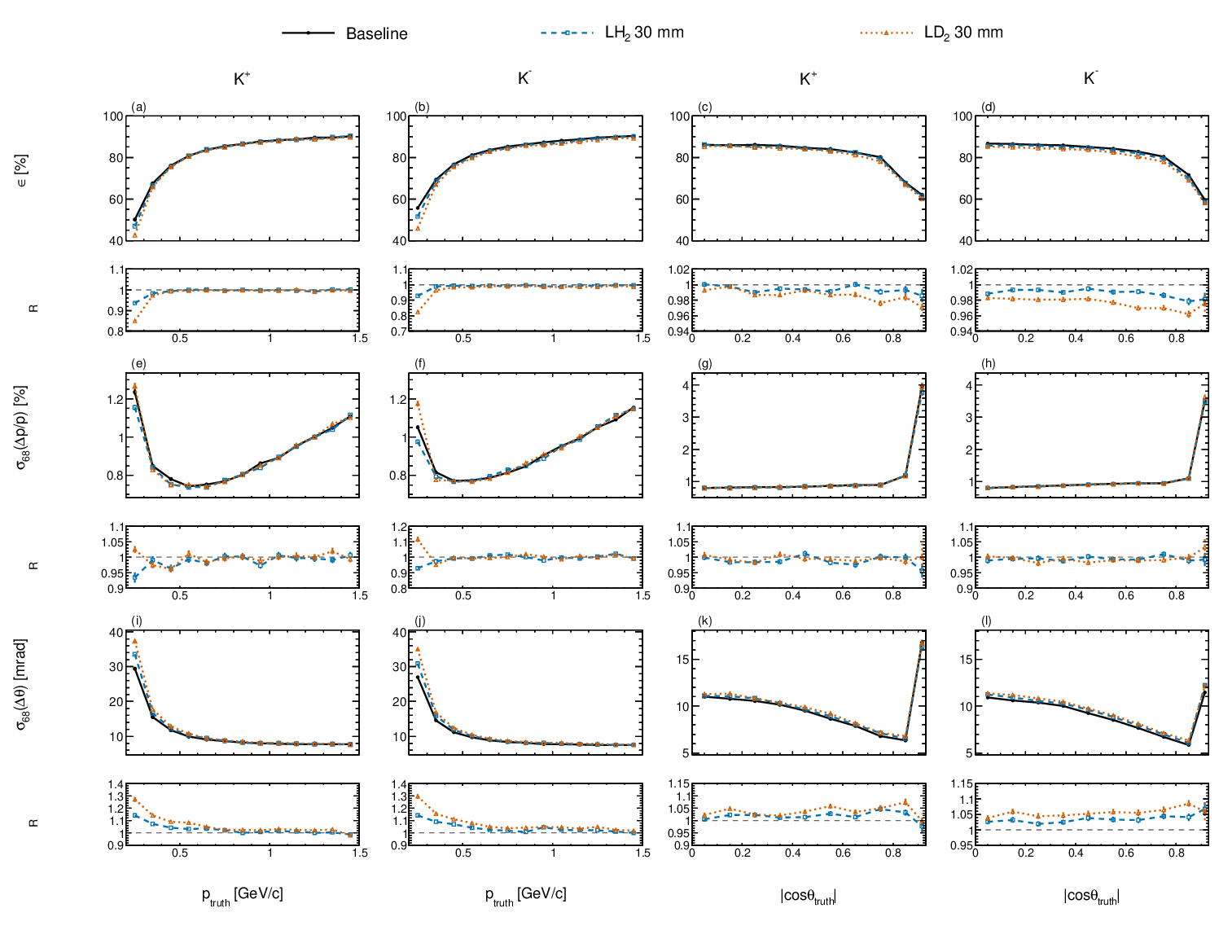}
  \caption{Single-particle tracking performance for $K^+$ and $K^-$, using the same observables, layout, configurations, and axis conventions as in Fig.~\ref{fig:single_particle_proton}. The truth-momentum range spans 0.2--1.5~GeV/$c$.}
  \label{fig:single_particle_kaon}
\end{figure*}

\begin{figure*}[t]
  \centering
  \includegraphics[width=\textwidth]{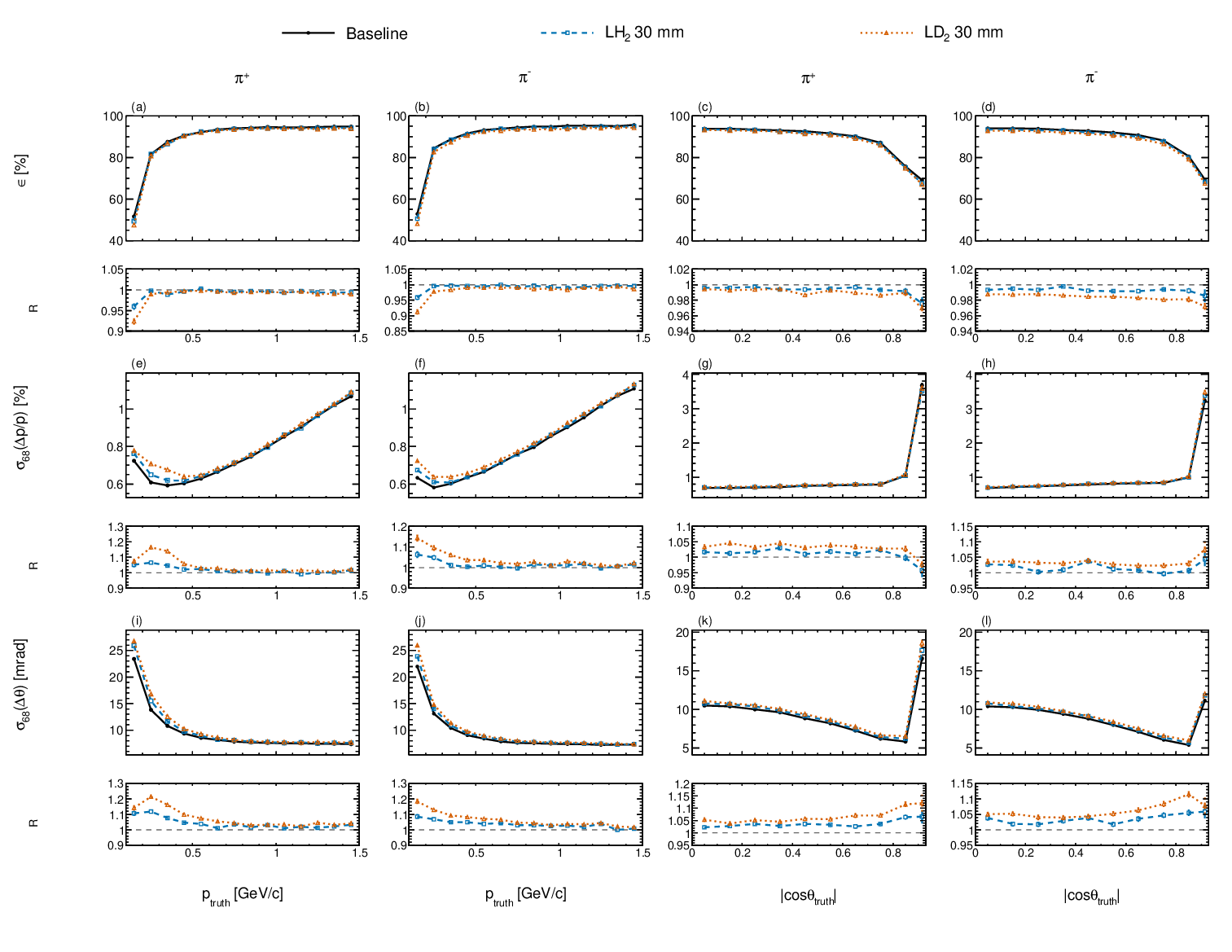}
  \caption{Single-particle tracking performance for $\pi^+$ and $\pi^-$, using the same observables, layout, configurations, and axis conventions as in Fig.~\ref{fig:single_particle_proton}. The truth-momentum range spans 0.1--1.5~GeV/$c$.}
  \label{fig:single_particle_pion}
\end{figure*}

Figures~\ref{fig:single_particle_proton}--\ref{fig:single_particle_pion} display the tracking performance across the baseline and target configurations. Target-induced degradation is predominantly confined to low momenta below $0.5$~GeV/$c$ (e.g., up to a 22 percentage-point reduction for protons in LD$_2$). Above $0.5$~GeV/$c$, tracking efficiencies for protons, kaons, and pions remain robust, staying within $1.6$ percentage points of the baseline. Antiprotons are an exception, showing a systematic deficit of up to $2.2$ points in LH$_2$ and $4.9$ points in LD$_2$; this loss scales directly with nucleon density and is governed by in-flight annihilation within the liquid target.

\subsection{Full-event simulation of \texorpdfstring{$\bar{n}p\to2\pi^{+}\pi^{-}$}{antineutron-proton annihilation}}
\label{subsec:nbarp_simulation}
To evaluate full-event reconstruction, we simulate the complete reaction sequence $J/\psi\to p\pi^{-}\bar{n}$ followed by the annihilation $\bar{n}p\to2\pi^{+}\pi^{-}$. The four-momenta of the tag proton, tag pion, and incident antineutron are sampled event-by-event using the PWA generator (Sec.~\ref{subsec:incident_beams}), while the annihilation products follow three-body phase space on a stationary target proton. This yields $98{,}661$ and $98{,}461$ simulated events in the beam pipe and LH$_2$ target, respectively.

Charged-track candidates are required to have valid hits in the MDC, successfully converged Kalman-filter fits, and polar angles within the tracking acceptance $|\cos\theta|<0.93$. The tag proton and negative pion must originate from the interaction region ($V_{xy}<1.0$~cm, $|V_z|<10.0$~cm) and are identified using combined $dE/dx$ and time-of-flight information. Each tag pair must pass a common vertex fit with $\chi^2\leq200$ and a one-constraint (1C) kinematic fit to the antineutron missing mass with $\chi^2\leq200$; if several pairs qualify, the one with the smallest sum of the two $\chi^2$ values is kept, and its recoil mass, computed from the vertex-fitted momenta, must lie within $[0.89, 0.99]$~GeV/$c^2$. Upon removing the tag pair, exactly three remaining charged tracks with net charge $+1$ ($++-$) are required. These three tracks are fitted to a common secondary vertex under the pion mass hypothesis, requiring $\chi^2<200$ and a transverse vertex radius $R_{xy}=\sqrt{x^2+y^2}\geq10$~mm.

Of the simulated events, $26{,}938$ in the beam pipe and $23{,}237$ in LH$_2$ pass all requirements. The conditional selection efficiency $\epsilon_{\mathrm{cond}}=N_{\mathrm{sel}}/N_{\mathrm{sim}}$ is $(27.30\pm0.14)\%$ for the beam pipe and $(23.60\pm0.14)\%$ for LH$_2$, giving a ratio of $0.864\pm0.007$. About half of the loss occurs at the requirement of exactly three signal tracks (54.1\% of tagged events for LH$_2$ versus 58.0\% for the beam pipe), with smaller losses in the tag selection (67.6\% versus 69.4\%) and in the signal-pion track fits (77.8\% versus 79.6\%).

For resolution studies, we evaluate the transverse radial residual,
\begin{equation}
\Delta r=r_{\mathrm{fit}}-r_{\mathrm{true}},\qquad r=\sqrt{x^2+y^2},
\label{eq:delta_r}
\end{equation}
where $r_{\mathrm{true}}$ is the transverse radius of the true annihilation vertex, and the energy closure,
\begin{equation}
\Delta E=\sum_{i=1}^{3}E_{\pi_i,\mathrm{VF}}-E_{\bar{n},\mathrm{tag\,1C}}-m_p,
\label{eq:delta_e}
\end{equation}
here, the pion energies $E_{\pi_i,\mathrm{VF}}$ are taken from the common three-pion vertex fit, and the antineutron energy $E_{\bar{n},\mathrm{tag\,1C}}$ is obtained from the tag-side 1C fit.

\begin{figure*}[t]
  \centering
  \subfloat[]{\includegraphics[width=0.48\textwidth]{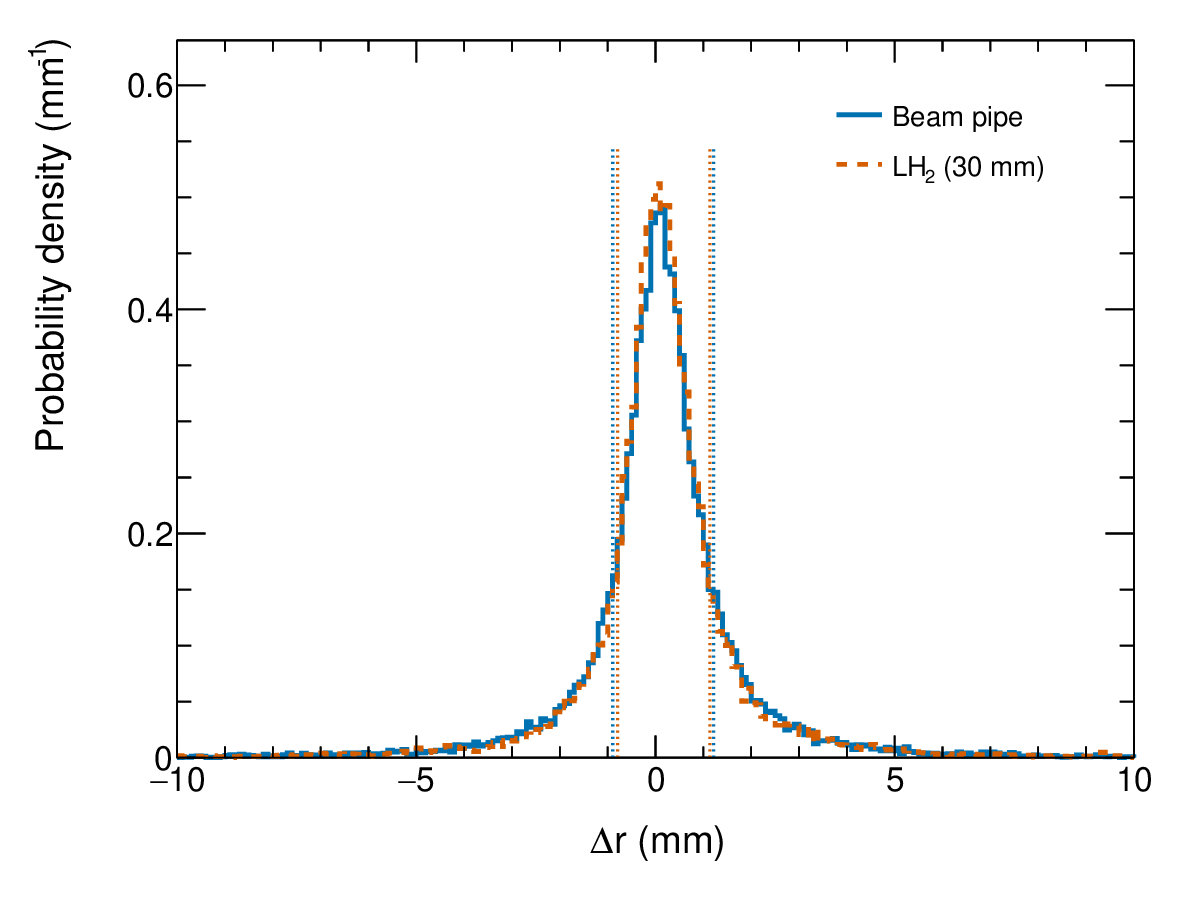}\label{fig:nbarp_delta_r}}\hfill
  \subfloat[]{\includegraphics[width=0.48\textwidth]{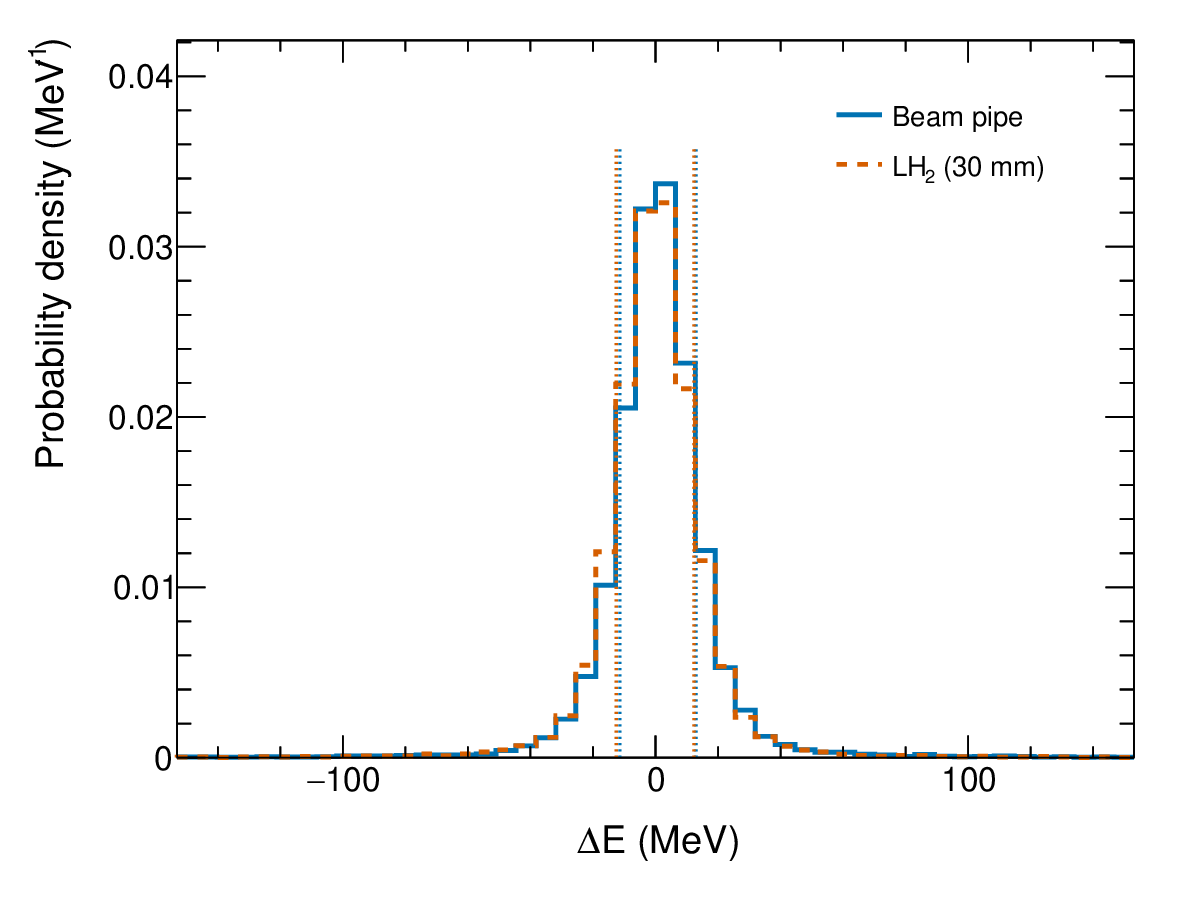}\label{fig:nbarp_delta_e}}
  \caption{Reconstruction performance for forced $\bar np\to2\pi^+\pi^-$ annihilations for truth-matched signal tracks in the beam-pipe and 30~mm LH$_2$ configurations: (a) radial residual $\Delta r$ of the common three-pion secondary vertex; (b) energy closure $\Delta E$ of the three-pion system relative to the tagged antineutron and a target proton at rest. Dashed lines mark the 16th and 84th percentiles of each full distribution. }
  \label{fig:nbarp_resolution}
\end{figure*}

Figure~\ref{fig:nbarp_resolution} compares the reconstructed distributions between the two target geometries. Resolution widths $\sigma_{68}$ are calculated as half the difference between the 84th and 16th percentiles of each distribution, with statistical errors determined via 2000 bootstrap resamples. The beam-pipe and LH$_2$ configurations exhibit radial widths of $1.054\pm0.010$ and $0.965\pm0.010$~mm, and energy-closure widths of $12.19\pm0.09$ and $12.43\pm0.10$~MeV, respectively. These results demonstrate that signal events originating in the cryogenic liquid are reconstructed with vertex and energy resolutions comparable to those in the beam pipe. The radial resolution scale also establishes the feasibility of cleanly isolating liquid interactions from beam-pipe backgrounds. The 6.7~mm physical gap between the outer beam-pipe surface and the inner liquid boundary corresponds to $6.4\,\sigma_{68}$ of the beam-pipe vertex resolution.

\section{Effective target densities and projected signal yields}
\label{sec:signal_yields}

\subsection{Effective target densities}
\label{subsec:target_densities}

To compare the cryogenic targets with the existing beam pipe (BP), we define the effective areal number density encountered by an incident particle of species $Y$ within $|\cos\theta|<0.8$ as
\begin{equation}
\mathcal{T}_Y = \frac{\rho \, \langle l\rangle_Y}{M}\, N_A,
\label{eq:L_target}
\end{equation}
where $\rho$ is the mass density, $M$ is the molar mass of the scattering center (atomic H for free protons in $\mathrm{LH}_2$; atomic D, providing one proton and one neutron, in $\mathrm{LD}_2$), and $N_A$ is Avogadro's number. We adopt $\rho = 0.0708~\mathrm{g\,cm^{-3}}$ and $M = 1.008~\mathrm{g\,mol^{-1}}$ for $\mathrm{LH}_2$, and $\rho = 0.164~\mathrm{g\,cm^{-3}}$ and $M = 2.014~\mathrm{g\,mol^{-1}}$ for $\mathrm{LD}_2$.

The effective path length $\langle l\rangle_Y$ is governed by the decay survival probability $S_Y(r,\theta)$ of Eq.~\eqref{eq:decay_integration}. A particle emitted at polar angle $\theta$ traverses a radial interval $dr$ along an oblique path element $dr/\sin\theta$; given azimuthal symmetry, only the polar-angle distribution $f_Y(\cos\theta)$ of Sec.~\ref{subsec:incident_beams} enters. For a cylindrical layer bounded by inner and outer radii $R_{\mathrm{in}}$ and $R_{\mathrm{out}}$, the mean path length of accepted particles is therefore
\begin{equation}
\langle l\rangle_Y = \frac{1}{A_Y} \int_{-0.8}^{0.8} f_Y(\cos\theta) \int_{R_{\mathrm{in}}}^{R_{\mathrm{out}}} \frac{S_Y(r,\theta)}{\sin\theta}\,dr\,d\cos\theta.
\label{eq:path_length_avg}
\end{equation}

For hyperons, the inputs are those of Table~\ref{tab:hyperon_decay_modes}. For antineutrons, which are stable on the scale of the detector ($S_{\bar n} = 1$), Eq.~\eqref{eq:path_length_avg} simplifies to $t\,\langle 1/\sin\theta\rangle$, where $t = R_{\mathrm{out}} - R_{\mathrm{in}}$ denotes the radial layer thickness. In this case, the average is evaluated over the 415,250 accepted PWA events that represent $f_{\bar n}$.

Applying Eq.~\eqref{eq:path_length_avg} to both the liquid layer and the beam-pipe cooling oil ($3.23 < r < 3.31~\mathrm{cm}$) yields the mean path lengths summarized in Table~\ref{tab:beam_paths}, with upstream material interactions neglected. Propagating the parameter uncertainties of Table~\ref{tab:hyperon_decay_modes} gives relative density uncertainties of 0.32\%--6.35\%, dominated by the hyperon lifetimes.

\begin{table*}[thbp]
  \centering
  \caption{Angular acceptance $A_Y$, conditional boundary survival probabilities $S_Y(R_{\mathrm{in/out}})$, and mean path lengths within $|\cos\theta_Y|<0.8$. Values for $\langle l\rangle_{\mathrm{liquid}}$ ($4.04 < r < 7.04$~cm) apply to both LH$_2$ and LD$_2$; the cooling oil spans $3.23 < r < 3.31$~cm. Hyperon uncertainties are propagated from Table~\ref{tab:hyperon_decay_modes}; antineutron uncertainties are MC statistical only.}
  \label{tab:beam_paths}
  \scriptsize
  \setlength{\tabcolsep}{3pt}
  \begin{tabular}{llllll}
  \toprule
  Species & $A_Y$ & $S_Y(R_{\mathrm{in}})$ & $S_Y(R_{\mathrm{out}})$ & $\langle l\rangle_{\mathrm{liquid}}$ (cm) & $\langle l\rangle_{\mathrm{oil}}$ (cm) \\
  \midrule
  $\Lambda$ & $0.76065\pm0.00027$ & $0.5360\pm0.0013$ & $0.3393\pm0.0014$ & $1.4857\pm0.0048$ & $0.05599\pm0.00011$ \\
  $\Sigma^{+}$ & $0.8587\pm0.0010$ & $0.10485\pm0.00076$ & $0.02058\pm0.00025$ & $0.1689\pm0.0016$ & $0.014175\pm0.000084$ \\
  $\Xi^{-}$ & $0.7529\pm0.0010$ & $0.2105\pm0.0030$ & $0.0682\pm0.0016$ & $0.4267\pm0.0079$ & $0.02586\pm0.00030$ \\
  $\Xi^{0}$ & $0.7579\pm0.0011$ & $0.419\pm0.011$ & $0.222\pm0.010$ & $1.066\pm0.038$ & $0.0457\pm0.0010$ \\
  $\Omega^{-}$ & $0.7787\pm0.0082$ & $0.01861\pm0.00098$ & $0.001098\pm0.000098$ & $0.0200\pm0.0013$ & $0.00344\pm0.00015$ \\
  $\bar n$ & $0.83050\pm0.00053$ & $1$ & $1$ & $3.44675\pm0.00078$ & $0.091913\pm0.000021$ \\
  \bottomrule
  \end{tabular}
\end{table*}

In the existing beam pipe (BP), free protons are present only in the 0.8~mm cooling-oil layer~\cite{BESIII:2009fln}, so the comparison of the LH$_2$ target with this baseline is model-independent. For the bound nucleons in Be, C, and Au, previous BESIII analyses~\cite{BESIII:2023trh,BESIII:2025bft,BESIII:2023clq} adopted a surface-reaction model, in which a strongly absorbed projectile interacts mainly near the nuclear surface and the cross section per nucleus scales as $A^{2/3}$ rather than $A$. Each nucleus then contributes $A^{2/3}Z/A$ effective protons and $A^{2/3}(A-Z)/A$ effective neutrons, normalized such that a free nucleon counts as one. For antineutrons, this scaling reproduces the measured annihilation cross sections on C to Pb at 50--400~MeV/$c$ within 2\%~\cite{Astrua:2002zg}. For hyperons, comparable nuclear data are scarce, and the effective number of nucleons need not be the same for every reaction: it depends on how strongly the projectile is absorbed and on whether the reaction products rescatter or are captured in the nucleus. The few available estimates suggest deviations of tens of percent. An eikonal calculation for $^{12}$C$(\Xi^-,\Lambda\Lambda)X$ gave 3.0--4.1 effective protons, depending on the assumed cross section and $\Lambda$ trapping probability~\cite{Ahn:2005jz}, 14--56\% above the surface value of 2.62, and the BESIII $\Sigma^+n$ and $\Xi^0n$ analyses~\cite{BESIII:2025bft,BESIII:2023clq} adopted about three effective neutrons for $^9$Be, 25\% above the surface value of 2.40. Such differences leave the order-of-magnitude comparisons in Tables~\ref{tab:target_properties} and~\ref{tab:besiii_examples}, which use the surface values, unchanged, but they limit the precision of beam-pipe measurements in a way that more statistics cannot overcome. We take the 25\% difference for $^9$Be as the size of this uncertainty, assign it to the bound-nucleon densities of the beam pipe in Table~\ref{tab:target_properties}, and list it in Table~\ref{tab:systematic_comparison}. The LD$_2$ target avoids it by replacing the composite Be, C, and Au nuclei with the weakly bound deuteron ($B=2.22$~MeV), for which the nuclear corrections reduce to a few-body problem.

Combining these target definitions with the effective path lengths evaluated in Table~\ref{tab:beam_paths}, we obtain the effective areal number densities summarized in Table~\ref{tab:target_properties}. The beam-pipe neutron and proton densities are the surface-model values, with the 25\% nuclear-model uncertainty given as a separate error. Quantitatively, the 30~mm $\mathrm{LH}_2$ target provides a dramatic enhancement in free-proton areal density, exceeding that of the cooling oil by factors of 7--22 for $\Lambda$, $\Sigma^+$, $\Xi$, and $\bar{n}$. This gain peaks at 21.7 for stable antineutrons and gradually decreases for shorter-lived hyperons (6.9 for $\Sigma^+$ and 3.4 for $\Omega^-$) due to decay losses prior to reaching the liquid volume. For neutrons, by contrast, $\mathrm{LD}_2$ yields 0.4--2.9 times the surface-weighted beam-pipe density. The primary merit of the $\mathrm{LD}_2$ target is therefore qualitative rather than purely statistical: it provides a comparable effective neutron areal density while replacing heavy-nucleus medium effects with those of the few-body deuteron system.

\begin{table*}[t]
  \caption{Effective areal number densities $\mathcal{T}_Y$ (in units of $10^{22}$~cm$^{-2}$) for the beam pipe (BP) and the 30~mm liquid targets, for incident particles within $|\cos\theta_Y|<0.8$. For the BP, the neutron ($n$) and total-proton ($p$) columns use the surface prescription described in the text, and the free-proton column ($p_f$) is the hydrogen in the cooling oil. The LH$_2$ column gives free protons; the LD$_2$ column counts one neutron or one proton per deuteron. The first error propagates the hyperon inputs, or the finite PWA statistics for $\bar n$. The BP $n$ and $p$ entries carry a second error, the 25\% nuclear-model uncertainty of the effective number of bound nucleons (Sec.~\ref{subsec:target_densities}). Material and geometry uncertainties are not included.}
  \label{tab:target_properties}
  \centering
  \scriptsize
  \setlength{\tabcolsep}{4pt}
  \begin{tabular}{llllll}
  \toprule
  Beam $Y$ & BP ($n$) & BP ($p$) & BP ($p_f$) & LH$_2$ ($p_f$) & LD$_2$ ($n$/$p$) \\
  \midrule
  $\Lambda$ & $3.5447\pm0.0069\pm0.89$ & $3.3294\pm0.0065\pm0.83$ & $0.40843\pm0.00080$ & $6.284\pm0.020$ & $7.286\pm0.024$ \\
  $\Sigma^{+}$ & $0.9042\pm0.0053\pm0.23$ & $0.8482\pm0.0050\pm0.21$ & $0.10340\pm0.00061$ & $0.7143\pm0.0066$ & $0.8281\pm0.0076$ \\
  $\Xi^{-}$ & $1.644\pm0.019\pm0.41$ & $1.543\pm0.018\pm0.39$ & $0.1886\pm0.0022$ & $1.805\pm0.033$ & $2.093\pm0.039$ \\
  $\Xi^{0}$ & $2.898\pm0.064\pm0.72$ & $2.721\pm0.060\pm0.68$ & $0.3335\pm0.0074$ & $4.51\pm0.16$ & $5.23\pm0.19$ \\
  $\Omega^{-}$ & $0.2215\pm0.0095\pm0.055$ & $0.2074\pm0.0089\pm0.052$ & $0.0251\pm0.0011$ & $0.0844\pm0.0054$ & $0.0979\pm0.0062$ \\
  $\bar n$ & $5.8030\pm0.0013\pm1.5$ & $5.4533\pm0.0012\pm1.4$ & $0.67048\pm0.00015$ & $14.5792\pm0.0033$ & $16.9023\pm0.0038$ \\
  \bottomrule
  \end{tabular}
\end{table*}

\subsection{Projected signal yields}
\label{subsec:yield_projections}

To project the experimental sensitivity, we define two reaction counts for each channel. The first is the number of reactions produced in the target prior to tag and signal reconstruction:
\begin{equation}
N_{\mathrm{prod}} = N_{\mathrm{parent}}\,\mathcal B_{\mathrm{tag}}\,A_Y\,\mathcal T_Y\,\sigma_N,
\label{eq:n_prod}
\end{equation}
where $N_{\mathrm{parent}}$ is the number of produced $J/\psi$ or $\psi(3686)$ events, $\mathcal B_{\mathrm{tag}}$ is the production branching fraction (Table~\ref{tab:hyperon_decay_modes}), and $\sigma_N$ is the reaction cross section per target nucleon, treated as constant over the incident momentum range. Free-proton reactions use measured $\sigma_{Yp}$ values. For $\Lambda p\to\Sigma^+X$ and the neutron reactions, $\sigma_N$ is a nucleon-level reference value inferred from Be: the published proton-level value for $\Lambda p\to\Sigma^+X$, obtained with the surface factor 1.93~\cite{BESIII:2023trh}, and the Be cross sections measured by BESIII~\cite{BESIII:2025bft,BESIII:2023clq} divided by the surface factor 2.404 for the neutron reactions.

The second count estimates the reconstructed signal yield by scaling the yield $N_{\mathrm{BP}}^{\mathrm{pub}}$ observed in the corresponding published beam-pipe (BP) analysis:
\begin{equation}
 N_{\mathrm{rec}}^{\mathrm{proj}}
 =N_{\mathrm{BP}}^{\mathrm{pub}}\,G_{\mathrm{parent}}\,G_{\mathrm{target}}, \qquad
 G_{\mathrm{parent}}=\frac{N_{\mathrm{parent}}}{N_{\mathrm{parent}}^{\mathrm{BP}}},\quad
 G_{\mathrm{target}}=\frac{\mathcal T_{Y,\mathrm{liquid}}}{\mathcal T_{Y,\mathrm{BP}}}.
 \label{eq:yield_gain_factors}
\end{equation}
When the published baseline measures the same reaction on the same kind of target nucleon, the cross section $\sigma_N$, the tagging branching fraction $\mathcal B_{\mathrm{tag}}$, and the absolute reconstruction efficiencies cancel. This projection assumes comparable overall efficiencies between the liquid target and the beam pipe ($\bar\epsilon_{Y,\mathrm{liquid}}\approx\bar\epsilon_{Y,\mathrm{BP}}$). We keep this assumption for all channels: $\bar np\to2\pi^+\pi^-$ is the only channel simulated with the full event (Sec.~\ref{subsec:nbarp_simulation}), its efficiency ratio of 0.864 is moderate, and the selection, which was taken over from the beam-pipe analysis, can be optimized for the liquid target.

To establish a defensible annual benchmark, we assume ten months (300 calendar days) of dedicated run time at either the $J/\psi$ or the $\psi(3686)$; the two samples below are alternative running scenarios, not a combined annual program. The projected parent event samples are extrapolated directly from the demonstrated daily luminosity rates of earlier resonance runs: $1200~\mathrm{pb}^{-1}$ accumulated over 86 days at the $J/\psi$ peak in 2019 (BESIII collaboration-meeting report, unpublished) and $3208.5~\mathrm{pb}^{-1}$ over 126 days at the $\psi(3686)$ in 2021~\cite{BESIII:NumberPsi3686}. Accounting for a factor of 1.3 gain in integrated luminosity enabled by top-up injection~\cite{BESIII:2020nme}, the expected annual delivered luminosities are $5.44~\mathrm{fb}^{-1}$ at the $J/\psi$ and $9.93~\mathrm{fb}^{-1}$ at the $\psi(3686)$. Using standard reference cross sections of $3400~\mathrm{nb}$ and $640~\mathrm{nb}$~\cite{Li:2006production}, these translate to annual planning benchmarks of $1.85\times10^{10}$ $J/\psi$ and $6.36\times10^9$ $\psi(3686)$ events. Because the published beam-pipe analyses utilized the historical dataset of $(1.0087\pm0.0044)\times10^{10}$ $J/\psi$ events~\cite{BESIII:NumberJpsi}, $G_{\mathrm{parent}}=1.834$ across all $J/\psi$ channels in Table~\ref{tab:besiii_examples}.

The inputs $\mathcal B_{\mathrm{tag}}$, $A_Y$, and $\mathcal T_Y$ are listed in Tables~\ref{tab:hyperon_decay_modes}, \ref{tab:beam_paths}, and~\ref{tab:target_properties}; Table~\ref{tab:besiii_examples} gives $G_{\mathrm{target}}$ for each channel together with the resulting projected counts. The total enhancement factor is the product $G_{\mathrm{parent}}G_{\mathrm{target}}$. The crab-waist (CW) column of Table~\ref{tab:besiii_examples} illustrates a hypothetical tenfold increase in annual exposure, as might be enabled by the crab-waist scheme~\cite{Zobov:2016sxm,Bogomyagkov:2016cw}, and equals $10N_{\mathrm{rec}}^{\mathrm{proj}}$. Uncertainties quoted on $N_{\mathrm{rec}}^{\mathrm{proj}}$ propagate the 7--13\% statistical errors of the published BP measurements, while those on $N_{\mathrm{prod}}$ propagate cross-section uncertainties.

For the free-proton channels $\Lambda p$, $\bar\Lambda p$, and $\bar np$, the LH$_2$ target increases annual yields by factors of 28--40 over published BP statistics, reaching $1.7\times10^3$ to $7.3\times10^3$ reconstructed events per year. For $\Lambda p\to\Sigma^+X$ and the quasi-free neutron reactions, Table~\ref{tab:besiii_examples} lists only the raw interaction counts $N_{\mathrm{prod}}$: $3.2\times10^4$ per year for $\Lambda p\to\Sigma^+X$ with LH$_2$, and $(1.7$--$2.6)\times10^3$ for $\Sigma^+n\to\Lambda p$ and $\Sigma^0p$ and $7.9\times10^3$ for $\Xi^0n\to\Xi^-p$ with LD$_2$. Reconstructed counts are not extrapolated because the published efficiencies do not transfer to these targets. For $\Lambda p\to\Sigma^+X$, Eq.~\eqref{eq:yield_gain_factors} does not apply~\cite{BESIII:2023trh}: the published sample also includes interactions in the MDC inner wall, which are not counted in $\mathcal T_{Y,\mathrm{BP}}$. For the neutron reactions, the overall efficiency depends strongly on Fermi motion, spectator momentum acceptance, and final-state interactions, which are not simulated here.

$\Omega^-$ scattering is not included in the projections: only 1.86\% of accepted $\Omega^-$ hyperons reach the liquid layer (Table~\ref{tab:beam_paths}). With $6.36\times10^9$ $\psi(3686)$ events, Eq.~\eqref{eq:n_prod} gives $N_{\mathrm{prod}}\approx0.23\,(\sigma_{\Omega p}/\mathrm{mb})$ per year; for an assumed $\sigma_{\Omega p}=20$~mb and an overall tag and signal efficiency of 10\%, this corresponds to about 0.5 reconstructed events per year, so $\Omega^-$--nucleon scattering remains out of reach at BESIII.

\begin{table*}[thbp]
  \centering
  \caption{Annual reference counts for the 30~mm liquid targets and $1.85\times10^{10}$ $J/\psi$ events ($G_{\mathrm{parent}}=1.834$); $N_{\mathrm{rec}}^{\mathrm{proj}}$ assumes unchanged efficiency. The density ratio $G_{\mathrm{target}}=\mathcal T_{Y,\mathrm{liquid}}/\mathcal T_{Y,\mathrm{BP}}$ uses the BP free-proton density for the free-proton channels, the total BP proton density for $\Lambda p\to\Sigma^+X$, and the BP neutron density for the neutron channels (Table~\ref{tab:target_properties}). For $\Lambda p\to\Sigma^+X$ and the neutron channels, $\sigma_N$ is a nucleon-level reference value inferred from Be, and only $N_{\mathrm{prod}}$ is given (Sec.~\ref{subsec:yield_projections}).  Published yields and cross sections are from Refs.~\cite{BESIII:2024geh,BESIII:2023trh,BESIII:2025yup,BESIII:2025bft,BESIII:2023clq}. The $\Lambda p$ and $\bar\Lambda p$ cross sections are those measured within $|\cos\theta^*|<0.9$~\cite{BESIII:2024geh}.  Errors on $N_{\mathrm{prod}}$ reflect cross-section statistical and systematic uncertainties; errors on $N_{\mathrm{rec}}^{\mathrm{proj}}$ propagate published BP statistical uncertainties only. The CW scenario illustrates an optional tenfold luminosity upgrade ($10N_{\mathrm{rec}}^{\mathrm{proj}}$).}
  \label{tab:besiii_examples}
  \scriptsize
  \setlength{\tabcolsep}{4pt}
  \renewcommand{\arraystretch}{1.12}
  \begin{tabular}{lrrrrrr}
    \toprule
    Process & $G_{\mathrm{target}}$ & $\sigma_N$ (mb) & $N_{\mathrm{prod}}$ ($10^3$) & $N_{\mathrm{BP}}^{\mathrm{pub}}$ & $N_{\mathrm{rec}}^{\mathrm{proj}}$ & CW scenario \\
    \midrule
    $\Lambda p_f\to\Lambda p$ & 15.39 & $12.2\pm1.6\pm1.1$ & $20.39\pm2.67\pm1.84$ & $60.9\pm7.8$ & $1719\pm220$ & $\sim17{,}000$ \\
    $\bar\Lambda p_f\to\bar\Lambda p$ & 15.39 & $17.5\pm2.1\pm1.6$ & $29.25\pm3.51\pm2.67$ & $72.0\pm8.5$ & $2032\pm240$ & $\sim20{,}000$ \\
    $\bar n p_f\to2\pi^+\pi^-$ & 21.74 & $2.7\pm0.3\pm0.3$ & $12.82\pm1.42\pm1.42$ & $59.0\pm7.7$ & $2353\pm307$ & $\sim24{,}000$ \\
    $\bar n p_f\to2\pi^+\pi^-\pi^0$ & 21.74 & $10.8\pm0.8\pm0.9$ & $51.29\pm3.80\pm4.27$ & $182.0\pm13.5$ & $7259\pm538$ & $\sim73{,}000$ \\
    $\bar n p_f\to2\pi^+\pi^-2\pi^0$ & 21.74 & $12.6\pm1.3\pm1.0$ & $59.84\pm6.17\pm4.75$ & $97.7\pm10.0$ & $3897\pm399$ & $\sim39{,}000$ \\
    \midrule
    $\Lambda p\to\Sigma^++X$ & 1.887 & $19.3\pm2.4\pm1.8$ & $32.26\pm4.01\pm3.01$ & $795\pm101$ & -- & -- \\
    \midrule
    $\Sigma^+n\to\Lambda p$ & 0.9158 & $18.8\pm5.0\pm3.0$ & $2.65\pm0.71\pm0.42$ & $77.6\pm20.8$ & -- & -- \\
    $\Sigma^+n\to\Sigma^0p$ & 0.9158 & $12.4\pm4.0\pm2.9$ & $1.75\pm0.57\pm0.40$ & $48.6\pm15.9$ & -- & -- \\
    $\Xi^0n\to\Xi^-p$ & 1.804 & $9.19\pm2.20\pm1.87$ & $7.88\pm1.89\pm1.61$ & $22.9\pm5.5$ & -- & -- \\
    \bottomrule
  \end{tabular}
\end{table*}

\subsection{Systematic uncertainties}
\label{subsec:systematic_uncertainties}

The liquid target changes the systematic uncertainties of the BP analyses used above in three ways: it removes none of the published sources for the free-proton channels, removes the terms that arise from the composite beam-pipe material for the nuclear-inferred and neutron channels, and adds sources of its own.

For the free-proton channels, the BP measurements already select interactions on the oil hydrogen, so LH$_2$ removes no published source. The 9.1\% systematic uncertainty of $\Lambda p$ and $\bar\Lambda p$ elastic scattering is dominated by tracking, PID, and branching-fraction terms, while the selection terms were already found to be negligible with a Barlow test~\cite{BESIII:2024geh}. The $\bar np$ analyses, in contrast, assign 0.5--4.5\% terms to $\pm10\%$ variations of the kinematic-fit, vertex-fit, scattering-position, and $P(p_{\mathrm{oil}})$ requirements and 0.2--1.4\% to the $\Delta E$ fit, which include statistical fluctuations of the 59--182 signal events, and 1.8--2.6\% to the intermediate $\rho$ and $\omega$ states~\cite{BESIII:2025yup}. These terms, 5.2--6.5\% in quadrature, could decrease with the 40 times larger sample, through Barlow tests of the selection variations and a better-constrained description of the intermediate states, although a larger sample may also reveal deviations that are not significant at present. In the optimistic case that these terms become negligible, about 5.2--5.6\% from the other sources would remain.

For $\Lambda p\to\Sigma^+X$, the published per-proton cross section was obtained from the $^9$Be measurement with a Be-to-proton cross-section ratio of 1.93 from a surface-reaction assumption, with no uncertainty assigned to this ratio (about 25\% from the effective-nucleon comparison of Sec.~\ref{subsec:target_densities}), and includes a 3.6\% term from an alternative nuclear scaling~\cite{BESIII:2023trh}; LH$_2$ removes both, as well as the 6.1\% term from the material of the cooling devices. For $\Sigma^+n$ and $\Xi^0n$, LD$_2$ removes the 7.2--7.3\% nuclear-ratio terms and the 6.5--12.8\% terms that the published analyses assign to simulating the neutron at rest~\cite{BESIII:2025bft,BESIII:2023clq}; these are replaced by deuteron corrections that remain to be quantified. The retained terms, 12.7--17.9\% for $\Sigma^+n$ and 16.2\% for $\Xi^0n$, still limit these channels, but they largely reflect the small published samples rather than the target. For example, the largest single term of the $\Sigma^+n\to\Sigma^0p$ measurement, 14.3\%, arises from simulating an isotropic angular distribution that a larger sample could measure~\cite{BESIII:2025bft}.

Operating with an internal liquid target introduces new systematic considerations across all channels. Interactions occurring in the beam pipe and the passive target layers now act as backgrounds, while tracking and particle-identification (PID) efficiencies must be recalibrated with data to account for the added material budget. Beam-pipe interactions can be cleanly separated from the liquid volume by reconstructed vertex radius, facilitated by the $6.4\,\sigma_{68}$ clearance demonstrated in Sec.~\ref{subsec:nbarp_simulation}, while residual backgrounds from the containment walls can be subtracted using dedicated empty-target runs normalized to the tagged incident flux. Regarding target normalization, uncertainties in the liquid density and the cold cell thickness replace those of the cooling-oil layer. The liquid density is governed by its operating temperature; a temperature stability of 0.01~K, as achieved by the MUSE target, corresponds to a temperature-induced density variation of only 0.02\%~\cite{Roy:2019add}; the absolute density normalization of the proposed target remains to be evaluated. The cold cell thickness is subject to thermal contraction and pressure-induced deflection, which must be experimentally calibrated for this annular geometry. Thin-walled cryogenic targets in comparable experiments report thickness uncertainties of 1--5\%~\cite{CLAS:2025inclusive,Ramazani-Sharifabadi:2021dd}; adding only a 5\% thickness uncertainty in quadrature to the published 9.1\% systematic uncertainty of the benchmark $\Lambda p$ measurement would give 10.4\%, before the other target-induced sources are included. A systematic breakdown across all channels is summarized in Table~\ref{tab:systematic_comparison}, categorizing sources into those removed, retained, or introduced by the liquid target, with unaffected terms combined in quadrature.
\begin{table*}[thbp]
  \centering
  \caption{Systematic uncertainties of the published BP cross-section measurements, grouped by their status with the 30~mm liquid target, and the sources added by the target. Sizes are relative uncertainties in percent, as published for $\Lambda p$ and $\bar\Lambda p$ elastic scattering~\cite{BESIII:2024geh}, $\Lambda N\to\Sigma^+X$~\cite{BESIII:2023trh}, $\Sigma^+n$~\cite{BESIII:2025bft}, $\Xi^0n$~\cite{BESIII:2023clq}, and the three $\bar np$ final states~\cite{BESIII:2025yup}. Removed sources are those specific to the composite beam-pipe nuclei and material; for LD$_2$, the nuclear terms are replaced by the deuteron corrections listed among the new sources. The effective number of bound nucleons was not assigned an uncertainty in the published analyses; the value given here is the 25\% difference between three and 2.40 effective $^9$Be neutrons (Sec.~\ref{subsec:target_densities}). Retained sources are given as the quadrature sum of all published terms not listed as removed. The new sources have not been evaluated numerically.}
  \label{tab:systematic_comparison}
  \renewcommand{\baselinestretch}{1}\footnotesize
  \setlength{\tabcolsep}{6pt}
  \begin{tabular}{@{}lll@{}}
    \toprule
    Status & Source & Published size (\%) \\
    \midrule
    \multirow{4}{*}{Removed} & Nuclear cross-section ratio & $\Sigma^+X$ 3.6; $\Sigma^+n$ 7.3; $\Xi^0n$ 7.2 \\
     & Effective number of bound nucleons & $\Sigma^+X$, $\Sigma^+n$, $\Xi^0n$ $\sim$25 (this work) \\
     & Fermi motion of bound neutrons & $\Sigma^+n$ 6.5, 12.8; $\Xi^0n$ 10.0 \\
     & Material of the cooling devices & $\Sigma^+X$ 6.1 \\
    \midrule
    Retained & All other sources (in quadrature) & $\Lambda p$, $\bar\Lambda p$ 9.1; $\bar np$ 7.5--8.4; $\Sigma^+X$ 6.3; $\Sigma^+n$ 12.7, 17.9; $\Xi^0n$ 16.2 \\
    \midrule
    \multirow{3}{*}{New} & Interactions in BP and walls & -- \\
     & Liquid density and cell thickness & -- \\
     & Deuteron corrections (LD$_2$ only) & -- \\
    \bottomrule
  \end{tabular}
\end{table*}

\section{Conclusion and discussion}
\label{sec:summary}

We have studied the feasibility of an internal LH$_2$ or LD$_2$ target between the beam pipe and the CGEM-IT at BESIII. The present beam-pipe measurements have two limitations: free protons exist only in the thin cooling-oil layer, which limits the statistics of free-proton scattering, and neutron scattering relies on neutrons bound in Be, C, and Au nuclei, which requires nuclear-model corrections. A pure hydrogen or deuterium target addresses both.

For the free-proton channels $\Lambda p$, $\bar\Lambda p$, and $\bar np$, a 30~mm LH$_2$ layer increases the effective free-proton areal density relative to the beam pipe by factors of 15.4 for the $\Lambda$ and $\bar\Lambda$ beams and 21.7 for the $\bar n$ beam. Scaled from the published beam-pipe yields at unchanged efficiency, one year of running would give 28--40 times the published samples, with statistical errors of 1.2--2.4\%, well below the published systematic uncertainties of 7.5--9.1\%. The gain for these channels is therefore statistical and benefits mainly differential measurements. For $\Lambda p\to\Sigma^+X$, whose published cross section was inferred from Be, LH$_2$ removes the nuclear-scaling (3.6\%) and cooling-material (6.1\%) terms and the uncertainty of the Be-to-proton ratio. For $\Sigma^+n$ and $\Xi^0n$, an LD$_2$ target, with neutron areal densities 0.92 and 1.80 times those of the beam pipe, replaces the composite beam-pipe nuclei by the deuteron. This removes the nuclear cross-section-ratio terms (7.2--7.3\%) and the bound-neutron Fermi-motion terms (6.5--12.8\%) of the published measurements, as well as the ambiguity of about 25\% in the effective number of Be neutrons, in exchange for deuteron corrections that remain to be quantified. For the $\Omega^-$, decays before the target leave about 0.5 reconstructed events per year for an assumed 20~mb cross section and 10\% efficiency, and about five even with the tenfold exposure of the illustrative crab-waist scenario, so $\Omega^-N$ scattering remains highly challenging at BESIII. None of these estimates includes channel-dependent efficiencies or backgrounds.

In the single-particle simulations, the target affects tracking mainly below 0.5~GeV/$c$. Above 0.5~GeV/$c$, the efficiencies of protons, kaons, and pions stay within 1.6 percentage points of the baseline, while the antiproton efficiency is lower by up to 2.2 (LH$_2$) and 4.9 (LD$_2$) points. In the signal-only full-event simulation of $\bar np\to2\pi^+\pi^-$, LH$_2$ lowers the conditional selection efficiency by 14\% relative to the beam pipe, and the vertex and energy-closure resolutions of liquid-target events, $\sigma_{68}\approx1.0$~mm and 12~MeV, are comparable to those of beam-pipe events. The 6.7~mm gap between the beam pipe and the liquid corresponds to 6.4 times the beam-pipe radial resolution, so interactions in the two can be separated by vertex radius.

The engineering of the target, including the heat load and cooling, installation clearances, compatibility with the beam vacuum and the CGEM-IT, and hydrogen safety, remains to be studied. If cryogenic operation proves impractical, room-temperature hydrogen- or deuterium-rich materials such as (CH$_2$)$_n$, (CD$_2$)$_n$, LiH, or D$_2$O could serve as alternatives, at the cost of additional nuclei in the target. Apart from the antineutron, the beams considered here come from two-body decays at fixed momenta; hyperons from multi-body $J/\psi$ and $\psi(3686)$ decays could extend the measurements over a continuous momentum range, and their yields and backgrounds remain to be evaluated. Overall, an internal hydrogen or deuterium target is a promising concept for higher-statistics and less model-dependent $YN$ and $\bar nN$ scattering measurements at BESIII and, with larger resonance samples, at future facilities such as STCF~\cite{STCF} and SCTF~\cite{SCTF}; it requires further engineering and channel-level validation.

\ifauthorversion
  \begin{acknowledgments}
This work is supported in part by the National Natural Science Foundation of China under contract No.~12361141819 and the Program of Science and Technology Development Plan of Jilin Province of China under contract No.~20230101021JC.
\end{acknowledgments}

\fi

\end{document}